\documentclass[showpacs,showkeys,amssymb,aps,twocolumn]{revtex4}

\usepackage{dcolumn}
\usepackage{graphicx}

\usepackage{amsmath}

\begin{document}

\title[Statistics of Baseline Calibration]{Statistical Aspects of Baseline Calibration
in Earth-Bound Optical Stellar Interferometry}

\author{Richard J. Mathar}
\homepage{http://www.strw.leidenuniv.nl/~mathar}
\email{mathar@strw.leidenuniv.nl}
\affiliation{Leiden Observatory, Leiden University, P.O. Box 9513, 2300 RA Leiden, The Netherlands}

\date{\today}

\begin{abstract}
Baseline calibration of a stellar interferometer is a prerequisite to data reduction
of astrometric operations. This technique of astrometry is triangulation of star positions. Since
angles
are deduced from the baseline and delay side of these triangles, length 
and pointing direction (in the celestial sphere) of the baseline vector at the
time of observation are key input data.
We assume that calibration follows from reverse astrometry;
a set of calibrator stars with well-known positions is observed and inaccuracies
in these positions are leveled by observing many of them for a common best fit.

The errors in baseline length and orientation angles
drop proportional to the inverse square roots of the number of independent data
taken, proportional to the errors in the individual snapshots of the delay, and proportional
to the errors in the apparent positions of the calibrators. Scheduling becomes
important if
the baseline components are reconstructed from
the sinusoidal delay of a single calibrator as a function of time.
\end{abstract}

\maketitle

\section{Overview}

Astrometry as realized by a contemporary optical stellar interferometer
is founded on the sensitivity to wave front tilts of stellar light,
which leads to a dephasing proportional to the tilt and proportional
to the distance between the two telescope's apertures. This is measured
by the amount of delay---in units of time or optical path length difference---added
by the observatory's optical infrastructure to the beam hitting
the telescope closer to the star to adjust the phase difference
of the light beams for close-to-coherent superposition at the detector.

We only address geometry  in this manuscript.
The wealth of non-statistical effects
of tidal Earth-crust motion, Earth axis drifts
\cite{McCarthyIERS32,CapitaineAAp406,CapitaineAAp432,Lambert03,LambertAA457,Kaplanarxiv06}, aberration
and general-relativistic wavefront tilts
which leave residual noise once the known quantities are accounted for
is left aside.

Mathematics of small differences demonstrates in Section \ref{sec:trig} how
measurement of the difference in the delay defined by an angular separation
of two ``science'' objects in the celestial sphere requests some knowledge
of the baseline vector---which we vaguely define as the separation between the two
input pupils of the two telescopes involved. Baseline calibration is the auxiliary
observation of well-known (in the astrometric sense) calibrator stars to deduce
this baseline geometry.

On can think of two pure forms. First there is a Fourier mode which matches
the sinusoidal delay as a function of time
with the three free parameters  of the baseline vector
tracking an individual star
(Section \ref{sec:istar}).
This minimizes the time overhead of slewing/pointing/acquisition/fringe-loop-lock
cycles. A characteristic sensitivity of the baseline
parameters means that is is favorable to reserve time slots six hours apart
\cite{SaundersAA455}.

This compares to the second form---which one may call field mode---in
which delays of a larger set of 
calibrator stars covering a wide range of altitudes and azimuths are gathered
(Section \ref{sec:field}).

The topic of this script is focused on the narrow question: given
statistical errors of the delay measurement in Fourier mode and statistical errors
in positions derived from star catalogs in the field mode, how long or how
many of them, respectively, does the calibration need to balance their effect to the
levels set by the astrometric mode.

\section{Fundamental Trigonometry}\label{sec:trig}
\subsection{Daily {OPD}}

We start with a summary of the fundamental geometry of observing
a star with perfectly stable telescopes orbiting
a fixed Earth axis without atmospheric or similar distortions \cite{MatharSAJ177}.

The terrestrial coordinate system defines geographic longitude $\lambda$,
latitude $\phi$ and altitude $H$ of two telescopes.
By conversion into a Cartesian frame and construction of
the mid-point between any pair of these, we define the geographic longitude $\lambda$
and geographic latitude $\phi$ of the baseline, which serves to define the
topocentric Alt-Az system for the baseline---which implies that $a_b=0$
since we define an individual system for each pair. In this
topocentric coordinate system, the baseline vector has length $b$
and splits into Cartesian coordinates
\begin{equation}
{\mathbf b} = b\left(
\begin{array}{cc}
-\cos A_b \cos a_b\\
\sin A_b \cos a_b\\
\sin a_b
\end{array}
\right)
\end{equation}
as a function of azimuth $A_b$ (South over West) and inclination $a_b$.
A rotation matrix $U$
\begin{equation}
U\equiv
\left(
\begin{array}{ccc}
-\sin\phi\cos\lambda & -\sin\phi\sin\lambda & \cos\phi \\
\sin\lambda & -\cos\lambda & 0\\
\cos\phi\cos\lambda & \cos\phi\sin\lambda & \sin\phi
\end{array}
\right)
\label{eq:U}
\end{equation}
transforms these coordinates to Cartesian
coordinates in a geocentric frame
for suitably defined celestial declination $\delta_b$ and hour angle $h_b$ \cite{MatharSAJ177},
\begin{equation}
{\mathbf b}=U\cdot b\left(
\begin{array}{cc}
\cos\delta_b \cos(\lambda-h_b)\\
\cos\delta_b \sin(\lambda-h_b)\\
\sin\delta_b
\end{array}
\right)
.
\end{equation}
Table \ref{tab.UT} illustrates this transformation for a Chilean
site. (The constant height $H$ above the ellipsoid implies
that neither telescopes nor baselines are co-planar.)

The direction to the star of right ascension $\alpha$ and declination $\delta$
at current azimuth $A$ and zenith angle $z$
is
\begin{equation}
{\bf s}=
\left( \begin{array}{c} -\cos A \sin z \\
\sin A \sin z \\
\cos z \\
 \end{array} \right)
=
U\cdot \left(
\begin{array}{c}
\cos\delta\cos(\lambda-h)\\
\cos\delta\sin(\lambda-h)\\
\sin\delta
\end{array}
\right)
.
\label{eq:sGeograph}
\end{equation}

\begin{table}
\caption{A model of telescope coordinates and the associated equatorial variables
of the Very Large Telescope Interferometer aligned with WGS84 conventions.
}
\begin{ruledtabular}
\begin{tabular}{cD{.}{.}{9}D{.}{.}{9}D{.}{.}{2}}
telescope &\multicolumn{1}{r}{$\lambda$ (rad)}&\multicolumn{1}{r}{$\phi$ (rad)}&
 \multicolumn{1}{c}{$H$ (m)} \\
\hline
U1 & -1.228800386 & -0.429833092 & 2635.43 \\
U2 & -1.228796107 & -0.429825122 & 2635.43 \\
U3 & -1.228790929 & -0.429819523 & 2635.43 \\
U4 & -1.228780856 & -0.429823005 & 2635.43 \\
\end{tabular}

\begin{tabular}{cD{.}{.}{1}D{.}{.}{4}D{.}{.}{4}D{.}{.}{1}D{.}{.}{1}D{.}{.}{1}}
baseline & \multicolumn{1}{c}{$b$} & \multicolumn{1}{c}{$\lambda$} & 
\multicolumn{1}{c}{$\phi$} & \multicolumn{1}{c}{$A_b$} &
  \multicolumn{1}{c}{$\delta_b$} & \multicolumn{1}{c}{$h_b$} \\
         & \multicolumn{1}{c}{(m)} & \multicolumn{1}{c}{(deg)} & 
\multicolumn{1}{c}{(deg)} & \multicolumn{1}{c}{(deg)} &
  \multicolumn{1}{c}{(deg)} & \multicolumn{1}{c}{(deg)} \\
\hline
U12 & 56.4  & -70.4050 & -24.8342 & -153.9 & 46.8 & -49.7 \\
U13 & 102.1 & -70.4048 & -24.6272 & -147.5 & 43.9 & -56.8 \\
U14 & 130.2 & -70.4045 & -24.6273 & -119.5 & 25.6 & -76.8 \\
U23 & 46.5  & -70.4047 & -24.6270 & -139.8 & 39.8 & -63.8 \\
U24 & 89.5  & -70.4044 & -24.6271 & -98.6  & 7.8  & -86.4 \\
U34 & 62.5  & -70.4042 & -24.6269 & -69.3  & -18.4& -99.0 \\
\end{tabular}
\end{ruledtabular}
\label{tab.UT}
\end{table}

The equation-of-motion of the optical path difference (OPD) $D$
is \cite{MatharSAJ177}
\begin{eqnarray}
D={\bf s}\cdot {\bf b}
&=&
b\left[\cos z \sin a_b+\sin z \cos a_b \cos(A-A_b)\right]
\nonumber
\\
&=&
b\left[\sin\delta \sin\delta_b+\cos\delta \cos \delta_b \cos(h-h_b)\right].
\label{eq:bfromD}
\end{eqnarray}
This is the common spherical coordinates formula for the angular distance
between two points, one fixed at $(\alpha,\delta)$, the other
at fixed $\delta_b$ cycling the polar axis in one sidereal day $T_d$,
at an angular velocity of
\begin{equation}
\omega_d=2\pi/T_d\approx 73\, \mu\mathrm{rad/s}.
\end{equation}

\begin{figure}
\includegraphics[width=8cm]{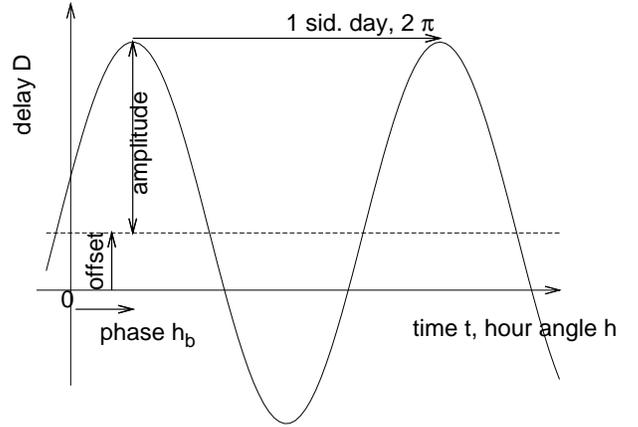}
\caption{
Equation (\ref{eq:bfromD}) generates a periodic delay in time.
}
\label{fig:sine}
\end{figure}

The characteristic  parameters of $D(h)$ are sketched in Figure \ref{fig:sine}:
a time-independent offset $b\sin\delta\sin\delta_b$, an amplitude $b\cos\delta\cos\delta_b$,
and a phase $h_b$.
The baseline length $b$ and the angles $\delta_b$ and $h_b$ are encoded in
phase and amplitude of $D$ plotted over time $t$.

The time-dependent projected baseline angle $p_b$ is
the position angle of high sensitivity of $D$ to changes
in the sky coordinates, that is the direction of high interferometric resolution
\cite{MatharSAJ177},
\begin{equation}
\tan p_b = \frac{\cos\delta_b\sin(h-h_b)}{\cos\delta\sin\delta_b-\sin\delta\cos\delta_b\cos(h-h_b)}
.
\label{eq:tanpb}
\end{equation}

\subsection{Differential {OPD}}

\begin{figure}
\includegraphics[width=8cm]{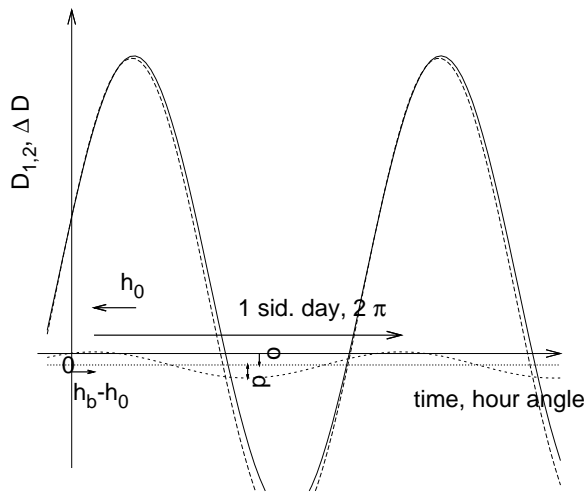}
\caption{
Equation (\ref{eq:fit24}) defines a differential delay with 
the same period  length as the wide-angle delays $D_1$ and $D_2$.
}
\label{fig:sineD}
\end{figure}

We are concerned with differential astrometry
of observing two objects at a single point in time
rather than switching between the two stars \cite{BellMess134,ArmstrongAJ496}.
We note (\ref{eq:bfromD}) for two stars with coordinates
$\delta_i$ and $\alpha_i=LST-h_i$,
mean positions $\bar\delta$ and $\bar\alpha$,
\begin{eqnarray}
\Delta \alpha&\equiv& \alpha_2-\alpha_1;
\quad
\bar \alpha\equiv (\alpha_2+\alpha_1)/2;
\\
\Delta \delta&\equiv& \delta_2-\delta_1;
\quad
\bar \delta\equiv (\delta_2+\delta_1)/2.
\\
\bar h &\equiv& (h_2+h_1)/2 = LST-\bar\alpha.
\end{eqnarray}
The time-independent cosine of the angular separation $\tau$ is
\begin{eqnarray}
\cos\tau&=&\cos\delta_1\cos\delta_2\cos(\alpha_1-\alpha_2)+\sin\delta_1\sin\delta_2 \\
& \approx & 1
-\frac{1}{2}(\Delta\delta)^2
-\frac{1}{2}\cos^2\bar\delta(\Delta\alpha)^2
+\frac{1}{24}(\Delta\delta)^4
\nonumber \\ &&
+\frac{1}{8}(\Delta\delta)^2(\Delta\alpha)^2
+\frac{1}{24}\cos^2\bar\delta(\Delta\alpha)^4
+\cdots
\end{eqnarray}
in these coordinates, neglecting sixth and higher order mixed
differentials.
The differential {OPD} becomes a
sinusoidal function of time, too,
\begin{equation}
\Delta D=D_2-D_1=({\bf s}_2-{\bf s}_1)\cdot {\bf b}
\equiv
o+p\cos(\bar h-h_b+h_0),
\label{eq:fit24}
\end{equation}
which defines a coordinate offset $o$,
a daily amplitude $p$, and a time shift $h_0$ (Fig.\ \ref{fig:sineD}).
For small $\Delta\alpha$ and small $\Delta\delta$,
up to fourth order in $\Delta\delta$,
\begin{eqnarray}
o
\approx
b\Delta\delta
\cos\bar\delta\sin\delta_b
-
\frac{1}{24}b(\Delta\delta)^3
\cos\bar\delta\sin\delta_b
.
\label{eq:o}
\end{eqnarray}
Up to fifth mixed order in the differentials we have the
squared amplitude of the differential delay,
\begin{eqnarray}
p^2 &\approx &
b^2\cos^2\delta_b\big[
\sin^2\bar\delta (\Delta \delta)^2
+
\cos^2\bar\delta (\Delta \alpha)^2
-
\frac{1}{12}\sin^2\bar\delta (\Delta \delta)^4
\nonumber
\\
&&
-
\frac{1}{12}\cos^2\bar\delta (\Delta \alpha)^4
-
\frac{1}{4} (\Delta \delta)^2 (\Delta \alpha)^2
\big]
\label{eq:psqr}
,
\end{eqnarray}
and the shift in hour angle
\begin{eqnarray}
\tan h_0 \approx
\cot\bar\delta \frac{\Delta\alpha}{\Delta\delta} 
\big[
1
-\frac{1}{12}(\Delta\delta)^2
+\frac{1}{12}(\Delta\alpha)^2
&&
\nonumber \\
-\frac{1}{720}(\Delta\delta)^4
-\frac{1}{144}(\Delta\delta)^2 (\Delta\alpha)^2
+\frac{1}{120}(\Delta\alpha)^4
\big]
.
&&
\label{eq:tanh0}
\end{eqnarray}
The right hand side contains a factor $\cos\bar\delta \Delta\alpha/\Delta\delta$,
the tangent of the position angle
of the binary to lowest order
in the differentials.

\subsection{Correlation of Variables}

The astronomer's interest of \emph{astrometric} operations is in the
measurement in the two parameters that characterize the relative
position of the binary's components on the sky, which could either
be represented as $\Delta\alpha$ and $\Delta\delta$, or alternatively
as distance $\tau$ and position angle.

From a fit to a sine model like (\ref{eq:fit24}), a measurement can
extract three parameters, which represent the six free Cartesian
components of head and tail of the baseline vector minus the
degrees of freedom of the three components of the baseline center (which represent
a free rigid
translation of the baseline and do not
contribute to the interferometric signal).

For the purpose of this script, the time base is not considered
an independent source of error. (On a real-time bus an
individual query for a time stamp has a resolution of roughly 10 $\mu$s,
i.e., $150$ $\mu$as after multiplication by $\omega_d$.
Micro-controller boards usually govern the readout process, so the jitter
is much smaller.)

These three parameters could either be stored as Cartesian coordinates
of the baseline vector,
or in the coordinates $b$, $\delta_b$
and $h_b$ which are more meaningful in the context of the analysis
of $D(t)$ in general.

Independent of this question of format, there is some redundancy
between the three fitting parameters $o$, $p$ and $h_0$ contained in $D(t)$
and the two positional parameters of the binary, supposed an independent baseline
calibration provides the auxiliary $b$,  $\delta_b$ and $h_b$.
The task of the baseline calibration is to support inversion of equations
(\ref{eq:o})--(\ref{eq:tanh0}). From this point of view, we need the
product $b\sin\delta_b$ to reduce (\ref{eq:o}), the product $b\cos\delta_b$
to reduce (\ref{eq:psqr}), and $h_b$ to reduce (\ref{eq:tanh0}).

If one of the three equations is not activated for some reason, one of
these three projections of the baseline vector (onto the polar axis and
on the equatorial plane) does not need to be calibrated either, because
---in principle---two equations for two unknowns remain. The U24 baseline
in Table \ref{tab.UT} provides an example of this aspect,
where $b\sin\delta_b\approx 12$ m.
The factor $\cos\bar\delta$ might reduce the product
$b\cos\bar\delta\sin\delta_b$ on the right hand side
of (\ref{eq:o}) to only $6$ m, which couples
an accuracy of 100 $\mu$as$=5\times 10^{-10}$ rad in $\Delta\delta$ to an
accuracy in $o$ of 3 nm. If the \emph{astrometric}
run
cannot meet
this requirement, equation (\ref{eq:o}) drops out of the data reduction
process, and in turn, the value $b\sin\delta_b$ is not requested from the
baseline calibration. If the value of $b\cos\delta_b$ is very small,
equation (\ref{eq:psqr}) may become disposable with the same  rationale.

From a similar mathematical but
entirely different engineering point of view,
the measurement of $D$ will likely
be assisted by a metrology system which is difficult to calibrate over
longer periods of time (which includes drifts of air densities if operating
in air and atmospheric lensing effects)
and is difficult to run without interruption.
The time derivative,
i.e, the delay velocities
\begin{eqnarray}
\Delta \dot D(t) = -p\omega_d \sin(\bar h-h_b+h_0),
\\
\dot D(t) = -b\omega_d \cos\delta\cos\delta_b \sin(h-h_b)
\label{eq:dotD}
\end{eqnarray}
will be available more easily, but this eliminates the offset, including
the differential offset $o$ and (\ref{eq:o}),
as an independent piece of information.

\subsection{Requested Baseline Accuracy}\label{sec:req}

The ``cushion'' and ``barrel'' distortions of the geometry
by the
cubic terms of (\ref{eq:o}) and biquadradic terms of (\ref{eq:psqr}) and (\ref{eq:tanh0})
are usually negligible: If the 
angular distance is limited by some finite field-of-view to $\tau\alt 2^\prime$,
the squares are limited to
$\tau^2\alt  3\times 10^{-7}$ rad$^2$. If the other terms in the sum are of the
order of unity, these higher order contributions change
$\Delta\delta$ or $\Delta\alpha$ by less than
$2^\prime\times 3\times 10^{-7}\approx 40$ $\mu$as.

So requirements on the baseline calibration can be
derived from the first orders
$o\sim b\Delta\delta \cos\bar\delta\sin\delta_b$
and $p\sim b\cos\delta_b\sin\bar\delta\Delta\delta$
$\sim b\cos\delta_b\cos\bar\delta\Delta\alpha$.
In a broad sense
the relative error in $\Delta\delta$
and $\Delta\alpha$ is of the order of the relative error in $o$ plus
the relative error in $b\sin\delta_b$, or $p$ plus the relative error
in $b\cos\delta_b$, respectively.
As a guideline: an accuracy of 100 $\mu$as in a field of
$\tau< 2^\prime$,
or 10 $\mu$as in a field of $\tau < 10^{\prime\prime}$,
is equivalent to a relative accuracy of $1\times 10^{-6}$,
which splits evenly into requirements of $5\times 10^{-7}$ in
$o$ ($p$) and $b$, which is a requirement of 50 $\mu$m for a baseline
of $b=100$ m of length.

Equivalent considerations are available for (\ref{eq:dotD}):
The relative error in the value of $b$ matches the relative error in
the value of $\dot D$.
For an error 50 $\mu$m in a 100 m baseline we need a
relative accuracy of $\dot D$ of $5\times 10^{-7}$.
Since the speed is of the order 1 cm/s for a 100 m baseline,
we want to have
$\dot D$ in absolute precision to $5\times 10^{-9}$ m/s = 5 nm/s.
Near the turning points $h\approx h_b$, the requirements for absolute
precision become tighter, enforced by the sine factor in (\ref{eq:dotD}).

The signal $\dot D(t)$ may also be small because either $\delta$
or $\delta_b$ are near 90$^\circ$, which puts higher stress on the
measurement of the \emph{absolute} value of this velocity. The effect through $\cos\delta$
means that a star on a closed apparent orbit around the corresponding celestial pole
of the observatory's hemisphere induces a small variation of delay because
its apparent position does not change much anyway. The effect through $\cos\delta_b$
punishes North-South baselines of observatories near the equator because
$\delta_b$ measures the angle between the baseline and the equatorial plane.

To keep the number of parameters basically down to three, this manuscript
does not introduce models of telescope axes runouts as a function of
azimuth and altitude.
They \emph{do} have an impact on the calibration: a
runout of 50 $\mu$m by one of the telescopes, i.e., not matched by the other,
is equivalent to that characteristic baseline tilt of 100 mas over $b=100$ m
mentioned above.

\section{Tracking a Single Star}\label{sec:istar}

\subsection{3-parametric fit}

The simplest scheme to derive baseline coordinates from reverse
wide-angle astrometry is to monitor the trace for a single star,
as in Fig.\ \ref{fig:sine}, over a significant portion of one of the
periods. 

One reason for such a strategy can be that there are few calibrator
stars with small errors in their $(\alpha,\delta)$ coordinates
and mixture
with data from more stars---strategy of Section \ref{sec:field}---could deteriorate the statistics.
Another reason is enhanced efficiency if the calibrator star is one
of the components of the binary system; baseline calibration and
astrometric observation may then run concurrently if delay and differential
delay data are recorded simultaneously.

A Fourier analysis can extract amplitude, offset and phase by a least-squares
fit
\begin{equation}
\sum_{j=1}^N [b\sin\delta\sin\delta_b+b\cos\delta\cos\delta_b\cos(h_j-h_b)-D_j]^2\rightarrow \mathrm{min}
\label{eq:fit24min}
\end{equation}
by $N$ measurements. The individual delay measurements $D_j$ are
characterized by an error variance $\sigma^2(D)$.
One might include some statistical weights of the squares in (\ref{eq:fit24min})
by some inverse function of the projected baseline---which represents the angular
resolution of the interferometer---but this adds more parameters to the present
analysis and obscures the principles laid out below.

The output (and free parameters) of the fitting process are
$b\sin\delta_b$ (polar component of the baseline),
$b\cos\delta_b$ (equatorial component of the baseline) and $h_b$
(interferometric hour angle). The error propagation of such a
Fourier analysis has been discussed in the literature
\cite[\S 7.04]{Smart2}\cite{RiceBSE23,GoldbergAO40}.
There is no need to split the components into length $b$
and angle $\delta_b$ for two reasons:
\begin{enumerate}
\item
These are correlated variables. There is no profit from dividing the
information (plus tracking the covariances), to re-unite them
afterwards for application in (\ref{eq:o}) and (\ref{eq:psqr}).
\item
If the the observable is the delay velocity $\dot D$ and the differential
delay velocity $\Delta \dot D$ as discussed above,
the components remain properly separated. One can basically ignore the
error analysis of the offsets (polar components).
\end{enumerate}

\subsection{Scheduling}

\begin{figure}
\includegraphics[width=8cm]{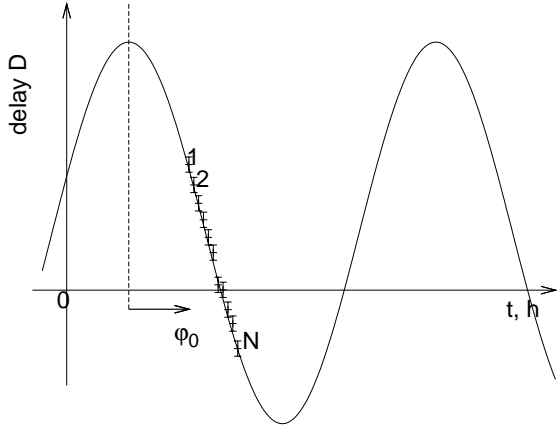}
\caption{
The baseline calibration in Fourier mode samples the delay
of Fig.\ \ref{fig:sine} at points $j=1,2,\ldots N$ in time, starting at $\varphi_0$.
}
\label{fig:sineN}
\end{figure}

The limits of observing during the night and observing
stars at some minimum altitude above the horizon lead to a characteristic
error propagation depending on which portion of the delay
is covered by the observation. As sketched in Fig.\ \ref{fig:sineN},
$N$ observations of snapshots $D_j$ would start at some time $\varphi_0$
after the daily maximum. The quality  of the fit to
the amplitude or offset depends on whether $\varphi$ is near zero or $\pi$
on one hand, or $\pi/2$ away from a maximum or minimum on the other hand.

\begin{figure}
\includegraphics[width=9cm]{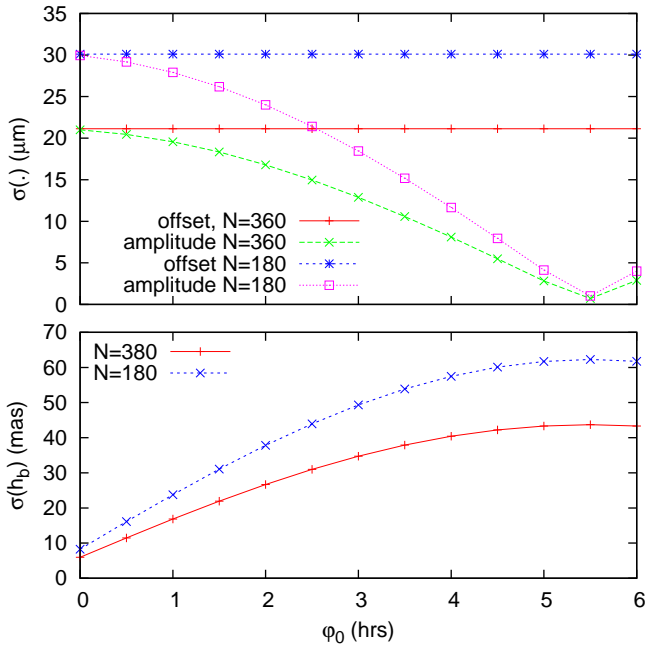}
\caption{
Errors to the delay offset, delay amplitude and delay hour angle for
$N=360$ measurements in
intervals of 10 seconds or $N=180$ measurements in intervals of 20 seconds
(an observation taking one hour).
$\sigma(D)=1$ $\mu$m for each measurement. 
}
\label{fig:vlbi}
\end{figure}

\begin{figure}
\includegraphics[width=8cm]{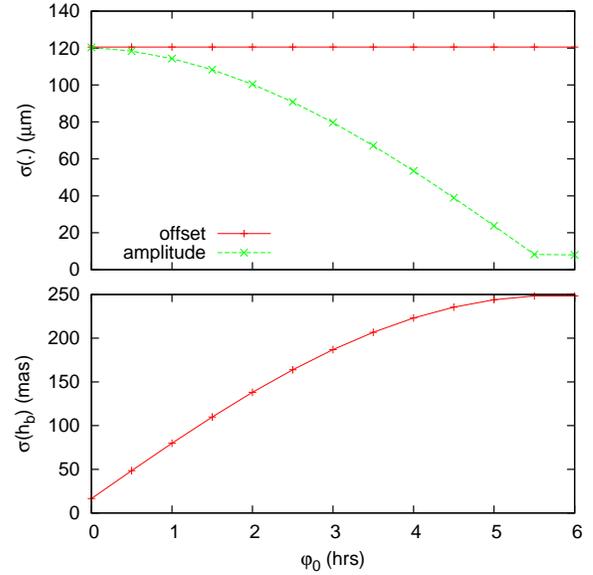}
\caption{
Errors to the three parameters of Fig.\ \ref{fig:vlbi} for $N=180$
measurements in intervals of 10 seconds, covering only the next half an hour after $\varphi_0$.
}
\label{fig:vlbi3}
\end{figure}

\begin{figure}
\includegraphics[width=9cm]{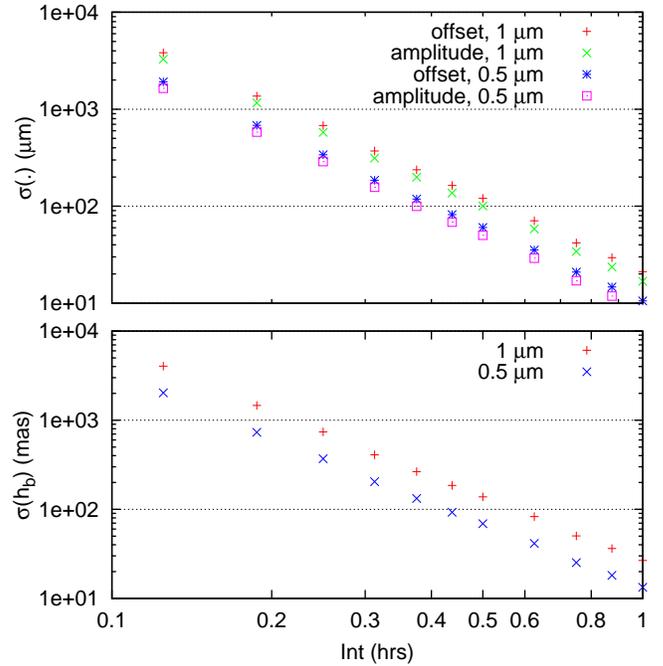}
\caption{
Errors to the three parameters for measurements at
intervals of 10 seconds, total integration time from 0.1 to 1 hour,
$\sigma(D)=1$ $\mu$m or 0.5 $\mu$m,
$\varphi_0=2$ hrs.
}
\label{fig:vlbi4}
\end{figure}

We summarize simple Monte Carlo calculations of these dependencies
in Figures \ref{fig:vlbi}--\ref{fig:vlbi4} for a baseline of $b=100$ m.
The common feature of Fig.\ \ref{fig:vlbi} and \ref{fig:vlbi3} is:
\begin{itemize}
\item
The quality of the measure of the offset (mean) does not depend on
when the observation is started.
\item
The amplitude of the curve is obtained with highest quality if the 
observation covers the part of maximum delay velocity. To achieve
highest precision in the hour angle, however, the observation ought to cover the part
of zero velocity near an extremum, six or eighteen hours earlier or later.
\end{itemize}
Fig.\ \ref{fig:vlbi}
shows 
the generic dependency of errors $\sim N^{-1/2}$ if duration and time
slot of the observation stay the same. The
transition from Fig.\ \ref{fig:vlbi} to Fig.\ \ref{fig:vlbi3}
demonstrates that cutting the observation time by half,
keeping the detector integration time the same,
increases the errors by an approximate factor of six. Fig.\ \ref{fig:vlbi4}
emphasizes this
dependence of the errors $\sim N^{-5/2}$ (at constant detector integration
time), but confirms the expected proportionality to the error $\sigma(D)$
of the individual  readout.

In this model, the error $\sigma(D)$ is an effective
superposition of an error induced by the error in the interferometric
phase plus an error from the jitter in the time base.
At velocities $\dot D<1$ cm/s, an error of $\sigma(h)=20$ $\mu$s
in the time base is equivalent to an error
$\sigma(D)=\dot D\sigma(h)< 200$ nm, for example.
Techniques to reduce this error by implementing detector-readout schedules
in low-level micro-controller programs are not
in the scope of this paper.

\section{Calibrator Star Catalog Imprecision}\label{sec:field}

\subsection{Baseline Length Calibration}
\begin{figure}
\includegraphics[width=9cm]{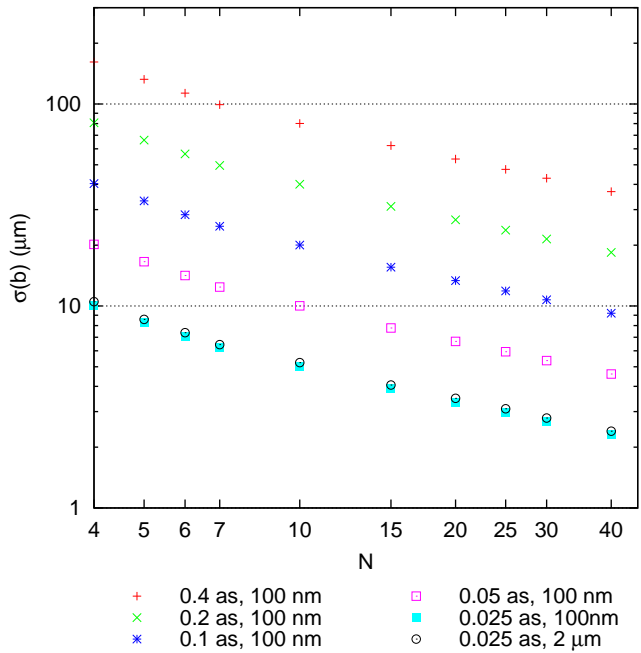}
\caption{
The error $\sigma(b)$ to a baseline length $b=100$ m from a least-squares fit to
$N$ measured delays $D$ for $N$ stars.
\label{fig:bcal}}
\end{figure}

The estimated error $\sigma(b)$ in the baseline length for $b=100$ m obtained by
a 1-parametric least squares fit after visiting $N$ stars
on the sky
that are rather homogeneously distributed in the range $z<60^\circ$
is shown in Figure \ref{fig:bcal}.
The $N$ positions are randomly selected
over the sky in the zenith range $z<60^\circ$,
taking subsets of the Hardin-Sloan-Smith points \cite{HardinIcover}, and
have been displaced by angles with five different Gaussian widths between
$0.025''$ and $0.4''$ to generate a measured delay, to which in addition
errors of $\sigma(D)=100$ nm or $\sigma(D)=2$ $\mu$m are added.
Double logarithmic axes scales are chosen to verify that
the reduction in $\sigma(b)$ is approximately proportional to $N^{-1/2}$.

\begin{figure}[hbt]
\includegraphics[width=9cm]{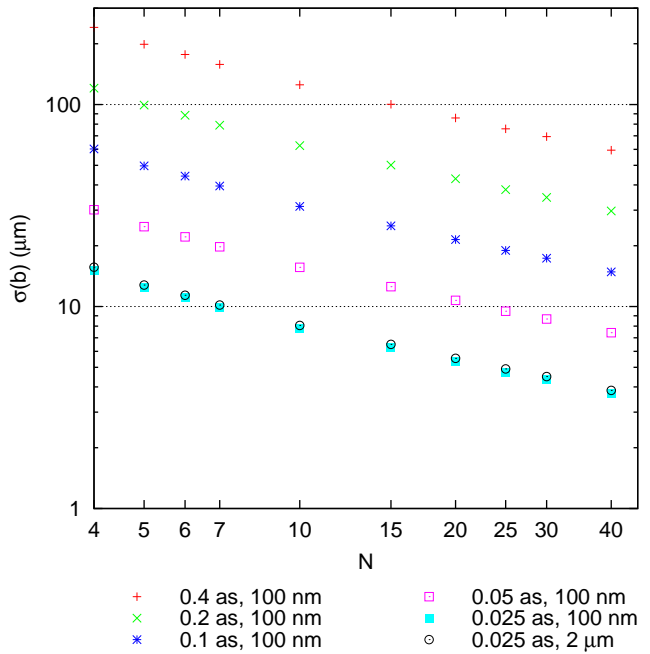}
\caption[Baseline calibration accuracy: constrained zenith]{
The errors of Figure \ref{fig:bcal} if baseline calibrator stars
are only selected in the zenith range $z<40^\circ$.
\label{fig:bcalz}}
\end{figure}

Figure \ref{fig:bcalz} uses a more constrained region
of the sky with $z<40^\circ$ and achieves inferior accuracy for equivalent
numbers of stars.
(The side effect that the projected baseline is larger on the average
than in Fig.\ \ref{fig:bcal} which implies better resolutions is not taken
into account.)

\begin{figure}
\includegraphics[width=9cm]{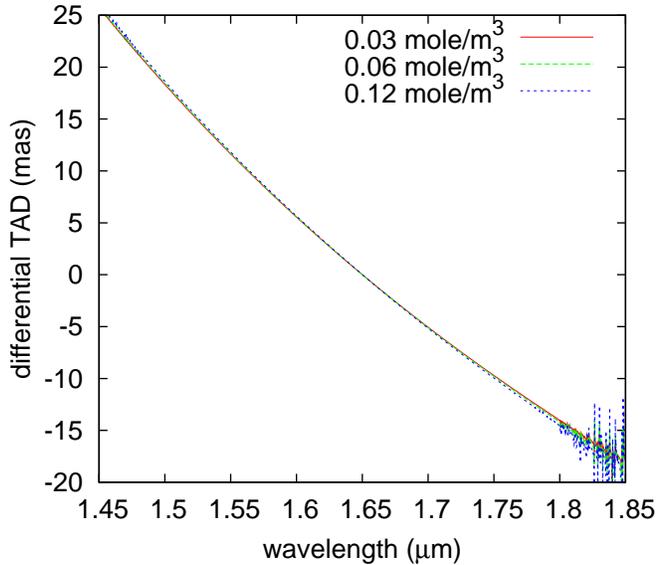}
\caption{The chromatic differential transverse atmospheric dispersion (TAD) in the H band
for $\tan z=1$ at an ambient pressure of 744 hPa, relative to a wavelength
of 1.65 $\mu$m.
}
\label{fig:tadH}
\end{figure}

\begin{figure}
\includegraphics[width=9cm]{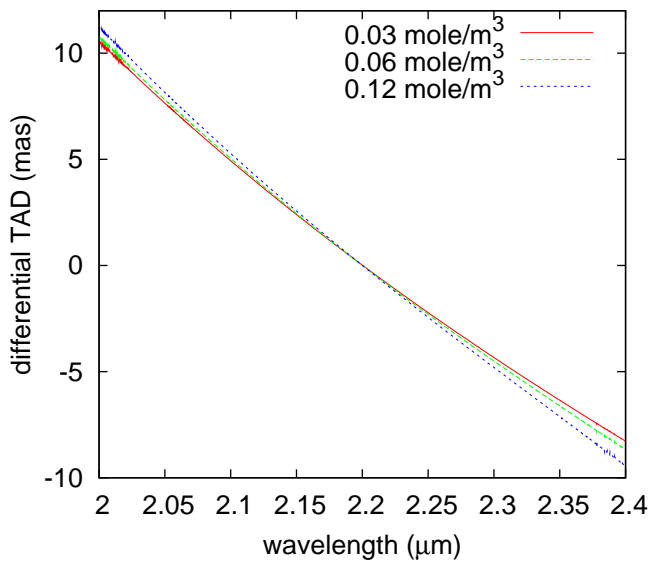}
\caption{The differential transverse atmospheric dispersion in the K band
for $\tan z=1$ at an ambient pressure of 744 hPa, relative to a wavelength
of 2.2 $\mu$m.
}
\label{fig:tadK}
\end{figure}

The apparent positions of stars are effected for example by the chromatic
dispersion in air, the product of the dielectric susceptibility
of air \cite{MatharJOptA9} at the telescope---measured in radians---and
the tangent of the zenith angle.
Examples for two infrared windows for Paranal conditions are plotted
in Fig.\ \ref{fig:tadH} and \ref{fig:tadK} for three different water
vapor densities. 

This implicit spread of a few tens of mas depending on the star
color is of the same order
as extrapolation errors for the apparent position 
from integrated proper motions
since the Hipparcos epoch \cite{DamlSAJ177,KovalevskyAA323}.

\subsection{Baseline Vector Calibration}

\begin{figure}[htb]
\includegraphics[scale=0.5]{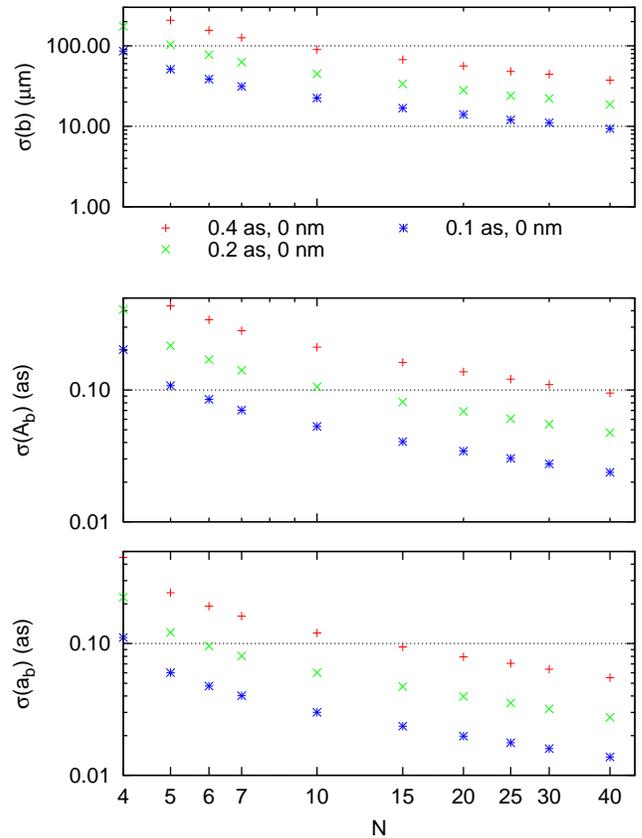}
\caption{
The error to a baseline length $b=100$ m, and to the baseline
orientation angles from a 3-parametric
least-squares fit measured to delays $D$ of $N$ calibrator stars.
}
\label{fig:bcal3}
\end{figure}

Fig.\ \ref{fig:bcal3} are results of a least squares fit of
the 3 degrees of freedom of the baseline vector to a series
of delay measurements,
minimizing $\sum_1^N [D_j-{\bf s}_j\cdot {\bf b}]^2$
over a set of ``noisy'' star positions
$A_j$ and $z_j$ and building a statistics over
the variables $b$, $A_b$ and $a_b$.
The error to the the measured delay has been set to zero, because
Figs.\ \ref{fig:bcal}--\ref{fig:bcalz} reveal that they are not important if
they remain $\sigma(D)\alt 2$ $\mu$m.
The $N$ positions have been rather homogeneously distributed over the sky
as in Fig.\ \ref{fig:bcal} in the zenith range $z<60^\circ$,
and have been randomly displaced by angles with different Gaussian widths
of $0.4''$ (pluses), $0.2''$ (crosses) or $0.1''$ (stars).

The accuracy in the baseline length $b$ is the
same as obtained with the 1-parameter fits of Fig.\ \ref{fig:bcal}\@.
Clearly, for small $N$, the error in the angles just echoes the error 
introduced in the star positions.

The elevation angle $a_b$ (tilt of the baseline versus
the horizontal) is obtained roughly twice as accurate
as the azimuth $A_b$. The interpretation of this bias is:
the even distribution
of the calibrator stars along azimuths and their more clumpy, overhead
distribution along zenith angles means that a measurement via
the projections $D$ achieves low resolution along the horizontal
for the two subsets of stars in the two opposite pointing
directions of the baseline \cite[Fig 2-8]{Muterspaugh05}.
The coupling along the vertical
coordinate is stiffer on the average, and the information
contained in the delays better distributed to deduce the baseline
tilt (pitch) versus the  horizon than the baseline rotation (roll) around the
zenith.

This difference in fitting quality in the $(a_b,A_b)$ angles
is more obscure in the $(\delta_b,h_b)$ system, because
multiplication with the inverse (transpose) of the $U$-matrix (\ref{eq:U})
weights components depending on the geographic latitude $\phi$.
Anyway, the angles in Fig.\ \ref{fig:bcal3}
are related to the effect of Earth axis pointing
\cite{KaplanIAUS141}, which can be written down as a change of the
effective ($\delta_b,h_b$) coordinates.
Formula (\ref{eq:bfromD}) is in fact symmetric as one could swap the variables
$\delta$ and $\delta_b$ or $h_j$ and $h_b$ without changing the delay.
Clearly, the calibration measures a baseline orientation in an ecliptic (celestial)
reference frame, not in a terrestrial reference frame. The two sides of this
coin are
\begin{itemize}
\item
Pushing errors in $a_b$ or $A_b$ below the limits set by the errors in the
IERS angles 
\texttt{www.usno.navy.mil/USNO/earth-orientation /eo-products}
is useless if the aim is to reduce delays to the terrestrial baseline data.
\item
There is a prospect of setting up a competitive optical reference frame
to 40 mas---aside from the influence of all systematic effects---if $N\agt 15$
stars with positions accurate to 100 mas are available.
(This estimate follows almost trivially from a $N^{-1/2}$ scaling of uncorrelated errors.)
\end{itemize}

A further remark: The baseline vector calibration is in a general sense equivalent
to determining the geographic longitude and latitude
of either head or tail of the baseline vector:
one can change the orientation of the baseline by
either tilting it explicitly or keeping it always horizontal
and sliding it with the tangent plane across the Earth surface.
In mathematical prose: Introducing the geographic latitude $\phi$ or longitude $\lambda$
as additional fitting parameters into the minimization procedure
defines an ill-conditioned problem.

\section{Summary} 

Baseline calibration reduces measurements of the scalar variable of the
delay to baseline vector
components along the polar
and equatorial axes, and splits the equatorial
component into two with the aid of clocks.

The interest is in the measure of angles, such that the requirements
on lengths (delay and baseline) are not formulated in absolute but in relative units: the relative error
in the star separation of the astrometric observation is the relative
error in the delay measure plus the relative error in the baseline length,
and similar generic statements apply separately to the components in
right ascension and declination.

The statistical errors in the baseline coordinates are proportional
to the errors of the individual delay (or its speed) and time stamp,
and proportional to the inverse square root of the number of independent
measurements, as expected. If the baseline coordinates are derived
from a Fourier analysis of the delay or delay speed as a daily
function of time, the statistical errors depend more decisively on schedules
and on the percentage of the sidereal period that is covered.

\acknowledgments{
This work was supported by the NWO VICI grant 639.043.201
``Optical Interferometry: A new Method for Studies of Extrasolar Planets.''
}

\appendix

\section{Tidal Motion} 

\begin{figure}[h]
\includegraphics[width=9cm]{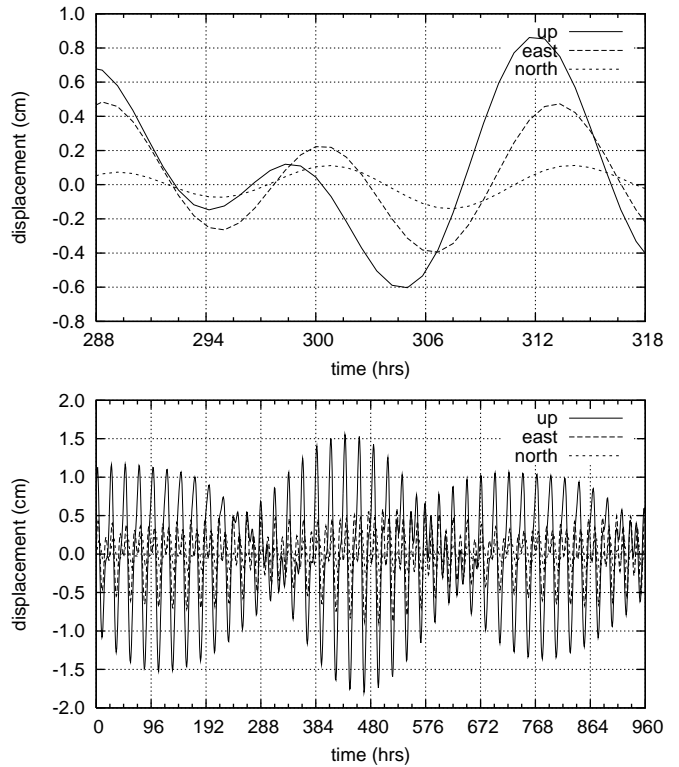}
\caption{Illustration of the vertical and lateral motion of the Earth crust at the Paranal
coordinates parametrized with 11 harmonic frequencies \cite{Bos}.
The fundamental mode exhibits the two daily tides, the envelope is
governed by the Moon's period.
\label{figure:earth_tide}}
\end{figure}

One contribution of the definition of the telescope coordinates in
an extra-terrestrial coordinate system is given by the influence by
the ocean tides that load and release the non-rigid Earth crust \cite{SoversRMP70}.
According to the \texttt{GOt00.2} model by Bos and Scherneck \cite{Bos},
the amplitude for the Paranal geographical coordinates is $<2$ cm
in vertical and $<0.7$ cm in horizontal directions: Fig.\ \ref{figure:earth_tide}.

The Earth crust motion in other regions of the planet may be up to 10 cm.

The periodic influence on astrometry can be estimated by converting the
East-West motion into  a change of the instantaneous geographic longitude
$\lambda$, the North-South motion into a change of the geographic latitude $\phi$.
A sliding by 1 cm translates into a tilt of $0.01/6.38\times 10^6\approx 1.5\times 10^{-9}$
rad $\approx 0.3$ mas, and is therefore not relevant for the baseline calibration
if the errors in calibrator star positions
are roughly one or two magnitudes larger.
With a similar  rationale, a change of $\bar\delta$ in (\ref{eq:o}) or (\ref{eq:psqr}) 
by $1.5\times 10^{-9}$ remains negligible if the request for relative
errors in $o$ or $p$ is of the order $5\times 10^{-7}$
as argued in section \ref{sec:req}.

\bibliographystyle{apsrmp}
\bibliography{all}

\begin{thebibliography}{21}
\expandafter\ifx\csname natexlab\endcsname\relax\def\natexlab#1{#1}\fi
\expandafter\ifx\csname bibnamefont\endcsname\relax
  \def\bibnamefont#1{#1}\fi
\expandafter\ifx\csname bibfnamefont\endcsname\relax
  \def\bibfnamefont#1{#1}\fi
\expandafter\ifx\csname citenamefont\endcsname\relax
  \def\citenamefont#1{#1}\fi
\expandafter\ifx\csname url\endcsname\relax
  \def\url#1{\texttt{#1}}\fi
\expandafter\ifx\csname urlprefix\endcsname\relax\def\urlprefix{URL }\fi
\providecommand{\bibinfo}[2]{#2}
\providecommand{\eprint}[2][]{\url{#2}}

\bibitem[{\citenamefont{Armstrong} \emph{et~al.}(1998)\citenamefont{Armstrong,
  Mozurkewich, Rickard, Hutter, Benson, Bowers, {Elias II}, Hummel, Johnston,
  Buscher, {Clark III}, Ha} \emph{et~al.}}]{ArmstrongAJ496}
\bibinfo{author}{\bibnamefont{Armstrong}, \bibfnamefont{J.}},
  \bibinfo{author}{\bibfnamefont{D.}~\bibnamefont{Mozurkewich}},
  \bibinfo{author}{\bibfnamefont{L.~J.} \bibnamefont{Rickard}},
  \bibinfo{author}{\bibfnamefont{D.~J.} \bibnamefont{Hutter}},
  \bibinfo{author}{\bibfnamefont{J.~A.} \bibnamefont{Benson}},
  \bibinfo{author}{\bibfnamefont{P.~F.} \bibnamefont{Bowers}},
  \bibinfo{author}{\bibfnamefont{N.~M.} \bibnamefont{{Elias II}}},
  \bibinfo{author}{\bibfnamefont{C.~A.} \bibnamefont{Hummel}},
  \bibinfo{author}{\bibfnamefont{K.~J.} \bibnamefont{Johnston}},
  \bibinfo{author}{\bibfnamefont{D.~F.} \bibnamefont{Buscher}},
  \bibinfo{author}{\bibfnamefont{J.~H.} \bibnamefont{{Clark III}}},
  \bibinfo{author}{\bibfnamefont{L.}~\bibnamefont{Ha}}, \emph{et~al.},
  \bibinfo{year}{1998}, \bibinfo{journal}{Astrophys.\ J.}
  \textbf{\bibinfo{volume}{496}}(\bibinfo{number}{1}), \bibinfo{pages}{550}.

\bibitem[{\citenamefont{Bos and Scherneck}(2005)}]{Bos}
\bibinfo{author}{\bibnamefont{Bos}, \bibfnamefont{M.~S.}}, and
  \bibinfo{author}{\bibfnamefont{H.-G.} \bibnamefont{Scherneck}},
  \bibinfo{year}{2005}, \bibinfo{title}{The free ocean tide loading provider},
  \urlprefix\url{http://www.oso.chalmers.se/~loading}.

\bibitem[{\citenamefont{Capitaine} \emph{et~al.}(2005)\citenamefont{Capitaine,
  Wallace, and Chapront}}]{CapitaineAAp432}
\bibinfo{author}{\bibnamefont{Capitaine}, \bibfnamefont{N.}},
  \bibinfo{author}{\bibfnamefont{P.~T.} \bibnamefont{Wallace}}, and
  \bibinfo{author}{\bibfnamefont{J.}~\bibnamefont{Chapront}},
  \bibinfo{year}{2005}, \bibinfo{journal}{Astron.\ Astrophys.}
  \textbf{\bibinfo{volume}{432}}(\bibinfo{number}{1}), \bibinfo{pages}{355}.

\bibitem[{\citenamefont{Capitaine} \emph{et~al.}(2003)\citenamefont{Capitaine,
  Wallace, and McCarthy}}]{CapitaineAAp406}
\bibinfo{author}{\bibnamefont{Capitaine}, \bibfnamefont{N.}},
  \bibinfo{author}{\bibfnamefont{P.~T.} \bibnamefont{Wallace}}, and
  \bibinfo{author}{\bibfnamefont{D.~D.} \bibnamefont{McCarthy}},
  \bibinfo{year}{2003}, \bibinfo{journal}{Astron.\ Astrophys.}
  \textbf{\bibinfo{volume}{406}}(\bibinfo{number}{3}), \bibinfo{pages}{1135}.

\bibitem[{\citenamefont{Damljanovi\'c and Pejovi\'c}(2008)}]{DamlSAJ177}
\bibinfo{author}{\bibnamefont{Damljanovi\'c}, \bibfnamefont{G.}}, and
  \bibinfo{author}{\bibfnamefont{N.}~\bibnamefont{Pejovi\'c}},
  \bibinfo{year}{2008}, \bibinfo{journal}{Serb.\ Astr.\ J.}
  \textbf{\bibinfo{volume}{177}}, \bibinfo{pages}{109}.

\bibitem[{\citenamefont{Goldberg and Bokor}(2001)}]{GoldbergAO40}
\bibinfo{author}{\bibnamefont{Goldberg}, \bibfnamefont{K.~A.}}, and
  \bibinfo{author}{\bibfnamefont{J.}~\bibnamefont{Bokor}},
  \bibinfo{year}{2001}, \bibinfo{journal}{Appl.\ Opt.}
  \textbf{\bibinfo{volume}{40}}(\bibinfo{number}{17}), \bibinfo{pages}{2886}.

\bibitem[{\citenamefont{Hardin} \emph{et~al.}(1997)\citenamefont{Hardin,
  Sloane, and Smith}}]{HardinIcover}
\bibinfo{author}{\bibnamefont{Hardin}, \bibfnamefont{R.~H.}},
  \bibinfo{author}{\bibfnamefont{N.~J.~A.} \bibnamefont{Sloane}}, and
  \bibinfo{author}{\bibfnamefont{W.~D.} \bibnamefont{Smith}},
  \bibinfo{year}{1997}, \bibinfo{title}{Tables of spherical codes},
  \urlprefix\url{http://www.research.att.com/~njas/packings/}.

\bibitem[{\citenamefont{Kaplan}(1990)}]{KaplanIAUS141}
\bibinfo{author}{\bibnamefont{Kaplan}, \bibfnamefont{G.~H.}},
  \bibinfo{year}{1990}, in \emph{\bibinfo{booktitle}{Inertial Coordinate
  Systems on the Sky}}, edited by \bibinfo{editor}{\bibfnamefont{J.~H.}
  \bibnamefont{Lieske}} and \bibinfo{editor}{\bibfnamefont{V.~K.}
  \bibnamefont{Abalakin}} (\bibinfo{publisher}{Kluwer},
  \bibinfo{address}{Dordrecht}), number \bibinfo{number}{141} in
  \bibinfo{series}{IAU Symposium}, pp. \bibinfo{pages}{241--250}.

\bibitem[{\citenamefont{Kaplan}(2006)}]{Kaplanarxiv06}
\bibinfo{author}{\bibnamefont{Kaplan}, \bibfnamefont{G.~H.}},
  \bibinfo{year}{2006}, \bibinfo{journal}{arXiv:astro-ph/0602086} .

\bibitem[{\citenamefont{Kovalevsky}
  \emph{et~al.}(1997)\citenamefont{Kovalevsky, Lindegren, Perryman, Hemenway,
  Johnston, Kislyuk, Lestrade, Morrison, Platais, R\"oser, Schilbach, Tucholke}
  \emph{et~al.}}]{KovalevskyAA323}
\bibinfo{author}{\bibnamefont{Kovalevsky}, \bibfnamefont{J.}},
  \bibinfo{author}{\bibfnamefont{L.}~\bibnamefont{Lindegren}},
  \bibinfo{author}{\bibfnamefont{M.~A.~C.} \bibnamefont{Perryman}},
  \bibinfo{author}{\bibfnamefont{P.~D.} \bibnamefont{Hemenway}},
  \bibinfo{author}{\bibfnamefont{K.~J.} \bibnamefont{Johnston}},
  \bibinfo{author}{\bibfnamefont{V.~S.} \bibnamefont{Kislyuk}},
  \bibinfo{author}{\bibfnamefont{J.~F.} \bibnamefont{Lestrade}},
  \bibinfo{author}{\bibfnamefont{L.~V.} \bibnamefont{Morrison}},
  \bibinfo{author}{\bibfnamefont{I.}~\bibnamefont{Platais}},
  \bibinfo{author}{\bibfnamefont{S.}~\bibnamefont{R\"oser}},
  \bibinfo{author}{\bibfnamefont{E.}~\bibnamefont{Schilbach}},
  \bibinfo{author}{\bibfnamefont{H.-J.} \bibnamefont{Tucholke}}, \emph{et~al.},
  \bibinfo{year}{1997}, \bibinfo{journal}{Astron. Astrophys.}
  \textbf{\bibinfo{volume}{323}}(\bibinfo{number}{2}), \bibinfo{pages}{620}.

\bibitem[{\citenamefont{Lambert}(2003)}]{Lambert03}
\bibinfo{author}{\bibnamefont{Lambert}, \bibfnamefont{S.}},
  \bibinfo{year}{2003}, \emph{\bibinfo{title}{Analyse et modelisation de haute
  precision pour l'orientation de la terre}}, Ph.D. thesis,
  \bibinfo{school}{L'Observatoire de Paris}.

\bibitem[{\citenamefont{Lambert}(2006)}]{LambertAA457}
\bibinfo{author}{\bibnamefont{Lambert}, \bibfnamefont{S.~B.}},
  \bibinfo{year}{2006}, \bibinfo{journal}{Astron.\ Astrophys.}
  \textbf{\bibinfo{volume}{457}}(\bibinfo{number}{2}), \bibinfo{pages}{717}.

\bibitem[{\citenamefont{Mathar}(2007)}]{MatharJOptA9}
\bibinfo{author}{\bibnamefont{Mathar}, \bibfnamefont{R.~J.}},
  \bibinfo{year}{2007}, \bibinfo{journal}{J. Opt.\ A: Pure and Appl. Optics}
  \textbf{\bibinfo{volume}{9}}(\bibinfo{number}{5}), \bibinfo{pages}{470}.

\bibitem[{\citenamefont{Mathar}(2008)}]{MatharSAJ177}
\bibinfo{author}{\bibnamefont{Mathar}, \bibfnamefont{R.~J.}},
  \bibinfo{year}{2008}, \bibinfo{journal}{Serb.\ Astr.\ J.}
  \textbf{\bibinfo{volume}{177}}, \bibinfo{pages}{115}, \bibinfo{note}{{E}: the
  sine in the equation on the second line of p 117 should be squared}.

\bibitem[{\citenamefont{McCarthy and Petit}(2003)}]{McCarthyIERS32}
\bibinfo{author}{\bibnamefont{McCarthy}, \bibfnamefont{D.~D.}}, and
  \bibinfo{author}{\bibfnamefont{G.}~\bibnamefont{Petit}},
  \bibinfo{year}{2003}, \emph{\bibinfo{title}{{IERS} Technical Note No 32}},
  \bibinfo{type}{Technical Report}, \bibinfo{institution}{{IERS} Convention
  Centre}, \urlprefix\url{http://www.iers.org/iers/publications/tn/tn32/}.

\bibitem[{\citenamefont{Muterspaugh}(2005)}]{Muterspaugh05}
\bibinfo{author}{\bibnamefont{Muterspaugh}, \bibfnamefont{M.~W.}},
  \bibinfo{year}{2005}, \emph{\bibinfo{title}{Binary Star Systems and
  Extrasolar Planets}}, Ph.D. thesis, \bibinfo{school}{Massachusetts Institute
  of Technology}.

\bibitem[{\citenamefont{Rice}(1954)}]{RiceBSE23}
\bibinfo{author}{\bibnamefont{Rice}, \bibfnamefont{S.~O.}},
  \bibinfo{year}{1954}, \emph{\bibinfo{title}{Mathematical Analysis of Random
  Noise}} (\bibinfo{publisher}{Dover}), pp. \bibinfo{pages}{133--294},
  \bibinfo{note}{reprinted from Bell System Journals 23, 24}.

\bibitem[{\citenamefont{Saunders} \emph{et~al.}(2006)\citenamefont{Saunders,
  Naylor, and Allan}}]{SaundersAA455}
\bibinfo{author}{\bibnamefont{Saunders}, \bibfnamefont{E.~S.}},
  \bibinfo{author}{\bibfnamefont{T.}~\bibnamefont{Naylor}}, and
  \bibinfo{author}{\bibfnamefont{A.}~\bibnamefont{Allan}},
  \bibinfo{year}{2006}, \bibinfo{journal}{Astron.\ Astrophys.}
  \textbf{\bibinfo{volume}{455}}(\bibinfo{number}{2}), \bibinfo{pages}{757}.

\bibitem[{\citenamefont{Smart}(1958)}]{Smart2}
\bibinfo{editor}{\bibnamefont{Smart}, \bibfnamefont{W.~M.}} (ed.),
  \bibinfo{year}{1958}, \emph{\bibinfo{title}{Combination of Observations}}
  (\bibinfo{publisher}{Cambridge University Press},
  \bibinfo{address}{Cambridge}).

\bibitem[{\citenamefont{Sovers} \emph{et~al.}(1998)\citenamefont{Sovers,
  Fanselow, and Jacobs}}]{SoversRMP70}
\bibinfo{author}{\bibnamefont{Sovers}, \bibfnamefont{O.~J.}},
  \bibinfo{author}{\bibfnamefont{J.~L.} \bibnamefont{Fanselow}}, and
  \bibinfo{author}{\bibfnamefont{C.~S.} \bibnamefont{Jacobs}},
  \bibinfo{year}{1998}, \bibinfo{journal}{Rev.\ Mod.\ Phys.}
  \textbf{\bibinfo{volume}{70}}(\bibinfo{number}{4}), \bibinfo{pages}{1393}.

\bibitem[{\citenamefont{{van Belle}} \emph{et~al.}(2008)\citenamefont{{van
  Belle}, {Sahlmann}, {Abuter}, {Accardo}, {Andolfato}, {Brillant}, {de Jong},
  {Derie}, Delplancke, {Duc}, {Dupuy}, {Gilli}} \emph{et~al.}}]{BellMess134}
\bibinfo{author}{\bibnamefont{{van Belle}}, \bibfnamefont{G.~T.}},
  \bibinfo{author}{\bibfnamefont{J.}~\bibnamefont{{Sahlmann}}},
  \bibinfo{author}{\bibfnamefont{R.}~\bibnamefont{{Abuter}}},
  \bibinfo{author}{\bibfnamefont{M.}~\bibnamefont{{Accardo}}},
  \bibinfo{author}{\bibfnamefont{L.}~\bibnamefont{{Andolfato}}},
  \bibinfo{author}{\bibfnamefont{S.}~\bibnamefont{{Brillant}}},
  \bibinfo{author}{\bibfnamefont{J.}~\bibnamefont{{de Jong}}},
  \bibinfo{author}{\bibfnamefont{F.}~\bibnamefont{{Derie}}},
  \bibinfo{author}{\bibfnamefont{F.}~\bibnamefont{Delplancke}},
  \bibinfo{author}{\bibfnamefont{T.~P.} \bibnamefont{{Duc}}},
  \bibinfo{author}{\bibfnamefont{C.}~\bibnamefont{{Dupuy}}},
  \bibinfo{author}{\bibfnamefont{B.}~\bibnamefont{{Gilli}}}, \emph{et~al.},
  \bibinfo{year}{2008}, \bibinfo{journal}{The Messenger}
  \textbf{\bibinfo{volume}{134}}, \bibinfo{pages}{6}.

\end{thebibliography}

\end{document}